\documentclass{JHEP3}

\usepackage{epsfig}
\usepackage{graphicx}

\title{Quantum Evolution of the Dilaton-Inflaton Model}

\author{V. N. Folomeev\\
    Physics
Institute of NAN KR, 265 a, Chui str., Bishkek, 720071,  Kyrgyz
Republic\\
    E-mail: \email{astra@freenet.kg}}

\abstract{The quantum evolution of a model of the universe
with account of two scalar fields ({\it dilaton} and {\it inflaton})
is considered. For this case, the closed and
flat models has been examined. It is shown that in both cases
the realization of conditions necessary for inflation is strongly
depends on coupling constant for dilaton and inflaton fields.
}
\begin{document}
\tableofcontents

\section{Introduction}

In Ref. \cite{ref:La} the new inflationary scenario (so-called extended inflation)
based on a combination of old
inflation \cite{ref:Guth} with Jordan-Brans-Dicke gravitational theory \cite{ref:Jord}
(hereafter JBD) has been offered. It allows to slow down a speed of
expansion of the universe at the initial stage of its evolution.
Thus there was power-law expansion of the universe  (instead of
exponential one). Such delay has allowed to solve a number of
problems of the old inflation noted still in Refs. \cite{ref:La} and \cite{ref:Wein}
(in particular, the problem of graceful exit). Unfortunately
in model of La and Steinhardt there was the basic difficulty
connected with impossibility of an establishment of thermal
equilibrium in walls of bubbles for cosmologically reasonable
times. It results in violation of isotropy of background
radiation. With the purpose of avoidance of this problem, it is
necessary that the coupling constant in JBD theory has been
limited from above $\omega \leq 20$  \cite{ref:Wein1}. However, the lower
observable bound is $\omega \geq 500$ \cite{ref:Reas}. For resolving of this
problem, Damour, Gibbons and Gundlach \cite{ref:Dam} (hereafter DGG) has
proposed the model in which, except of visible matter, the invisible
(field) component is present as well. Action in their model has a
form (in Jordan (physical) frame):
\begin{equation}
\label{actJBD}
S[g_{\mu \nu},\Phi,\psi_V,\psi_I]=S_{BD}[g_{\mu \nu},\Phi]+S_V[g_{\mu \nu},\psi_V]+
S_I[g_{\mu \nu},\Phi,\psi_I],
\end{equation}
where $\Phi$  is JBD scalar field ({\it dilaton}), $S_{BD}[g_{\mu \nu},\Phi]$ - Brans-Dicke
action \cite{ref:Jord}
\begin{equation}
\label{actBD}
S_{BD}[g_{\mu \nu},\Phi]=\int d^4 x \sqrt{-g}\left[-\Phi R+\frac{\omega}{\Phi}
g^{\mu\nu}\partial_{\mu}\Phi \partial_{\nu}\Phi\right],
\end{equation}
and $S_V$, $S_I$ are actions for visible $\psi_V$ and invisible
$\psi_I$ components of material fields. As the usual matter is described as
an ideal fluid, it does not play any role during inflation at this
stage of evolution of the universe and it is possible to dropped
it. Invisible component in DGG model reads:
\begin{equation}
\label{actI}
S_{I}[g_{\mu \nu},\Phi,\psi_I]=\int d^4 x \sqrt{-g}\left[\Phi^{1-\beta}g^{\mu\nu}
\partial_{\mu}\psi_I \partial_{\nu}\psi_I - \Phi^{2(1-\beta)}V(\psi_I)\right],
\end{equation}
where $\beta\equiv\beta_I/\beta_V$ and $\beta_{V,I}$ are
coupling constants of visible and invisible matter with dilaton
field. (Hereafter we work in geometrical units, i.e. $c=1$ and $8\pi G_N=1$, where $G_N$  is
the current value of Newtonian constant.) DGG worked in so-called
Einstein frame which turns out from Eq. (\ref{actJBD}) via conformal
transformation of the metrics $\tilde{g}_{\mu\nu}=2\Phi g_{\mu\nu}$ with simultaneous redefinition of
the variable $\Phi=\frac{1}{2} \exp{(\phi/\phi_0)}$, where $\phi_0\equiv\sqrt{\omega+3/2}$.
Using Eq. (\ref{actJBD}) Holman, Kolb and Wang \cite{ref:Holm} has
shown that at a choice of inflaton field as an invisible matter, it
is possible to find region on a plane $\omega - \beta$  in which the mentioned
above restrictions $\omega \leq 20$ (necessary for successful inflation) and $\omega \geq 500$ (claimed from
observations) can be satisfied simultaneously. It is possible
because,  instead of standard restriction $\omega \leq 20$ from model \cite{ref:La}, the new
relation $\omega/\beta^2\leq 20$ will work. It can be easily satisfied at $\omega \geq 500$.

The further development of the model was reduced to introduction in Eq. (\ref{actI}) arbitrary exponents $n$ and $m$
instead of $(1-\beta)$ and $2(1-\beta)$  \cite{ref:Holm1} that has allowed to investigate the generalized process
of nucleation and percolation and to calculate time-depended nucleation rate per unit volume.
In Ref. \cite{ref:Wang} the restrictions on parameters $n$ and $m$ at which the sufficient inflation was provided have been found.

It seems interesting to consider quantum evolution for the model of extended inflation. The main purpose of this paper is
to find necessary conditions for ensuring of inflation on post-quantum stage. In Section 2 the Wheeler-DeWitt equation
for closed, open and flat model is obtained in general form. Section 3 is devoted to problem statement and to
general description of solution technique. The derivation and solving of corresponding equations for closed and
flat models are presented in Section 4. Finally, in Section 5 the probabilities of tunneling and above the barrier reflection
for closed and flat models accordingly are found.

\section {Wheeler-DeWitt Equation}

\bigskip
We will work in Einstein frame. Thus, using conformal transformation mentioned
in Introduction and using designation $b=1-\beta$, we can obtain from  Eqs. (\ref{actBD}) and
(\ref{actI}) the following Lagrangian (see also Ref. \cite {ref:Bous}):
\begin {equation}
\label {lagr}
\mathcal{L} = \frac {1} {2} \sqrt {-\tilde{g}}[-\tilde{R}+\tilde{g}^{\mu \nu} \partial _ {\mu} \phi \partial _ {\nu} \phi +
e ^ {b\phi} \tilde{g}^{\mu \nu} \partial _ {\mu} \sigma \partial _ {\nu} \sigma -
e ^ {2b\phi} m^2 \sigma^2].
\end {equation}
Here the massive inflaton field $\sigma$ with the simplest form of
potential $V=\frac{1}{2} m^2 \sigma^2$ is using as invisible matter.
Let us write out the general expression for metrics for closed, flat and open universes:
\begin {equation}
\label {metr}
ds^2= N^2 (t) dt^2-a^2 (t) \left[ d\chi^2 + \left(\frac{\sin\sqrt{k}\chi}{\sqrt{k}} \right)
^2 (d\theta^2 +\sin^2\theta d\varphi^2) \right].
\end {equation}
In this expression $k =-1,0,1 $ are for closed, flat and open models
accordingly, $N (t)$ is lapse function. As is known,
the metrics in itself does not yet defines global topology of space-time.
It enables to consider spacelike  sections as
compact in a case of open and flat models. Thus the volume, limited
by spacelike hypersurfaces, remains finite and have a form:
\begin {equation}
\label {vol}
V_k =\int {d^3 x \sqrt {-h}}.
\end {equation}
($h$ is the three-dimensional metrics.) In a case of homogeneous model, using Eq. (\ref {metr}),
we can obtain:
\begin {equation}
\label {vol1}
V_k=a^3 v_k, \quad v_k =\int {d^3 x \frac{\sin^2\sqrt{k}\chi}{k}\sin{\theta}},
\end {equation}
where the value of $v_k$ depends on topology of the spacelike sections
(see, e.g., Ref. \cite{ref:Coule} and references inside).
Then, using Eqs. (\ref {lagr}) and (\ref {metr}), we have the following
Lagrange function:
\begin {equation}
\label {lagr1}
L = 3 v_k N \left [k a-\frac {1} {N^2} a \dot {a} ^2\right] +
\frac {1} {2} v_k N a^3\left [\frac {1} {N^2} \dot {\phi} ^2 +
\frac {e ^ {b \phi}} {N^2} \dot {\sigma} ^2-e^{2 b \phi} m^2 \sigma^2\right].
\end {equation}
Further, using Eq. (\ref {lagr1}) and calculating the conjugate momenta
\begin {equation}
\label {momen}
p_a =-\frac {6 v_k} {N} a \dot {a}, \quad
p_{\phi} = \frac {v_k} {N} a^3 \dot {\phi}, \quad
p_{\sigma} = \frac {v_k} {N} a^3 e^{b\phi}\dot {\sigma},
\end {equation}
we can obtain the corresponding canonical Hamiltonian:
\begin {equation}
\label {ham}
H=\frac{N}{v_k}\left [-\frac {1}{12}\frac {p_{a}^2}{a} +
\frac {1} {2 a^3} \left (p_{\phi}^2+e^{-b\phi} p_{\sigma}^2\right)-
3 v_k^2 a\left(k-\frac{1}{6} m^2 a^2 e^{2 b\phi} \sigma^2\right) \right].
\end {equation}
Quantization of Eq. (\ref {ham}) is reduced to replacement of the conjugate momenta
$p_a, p _ {\phi}, p _ {\sigma} $ on operators $-i \partial/\partial a $,
$-i \partial/\partial \phi $ and $-i \partial/\partial \sigma $
accordingly. Wheeler-DeWitt equation turns out at action
of the Hamilton operator on the wave function $ \psi (a, \phi, \sigma) $
and looks like:
\begin {equation}
\label {wd0}
-\frac{1}{12}a^{-p}\frac{\partial} {\partial a} \left (a^p \frac {\partial \psi} {\partial a} \right)
+ \frac {1} {2 a^2}
\left (\frac {\partial^2\psi} {\partial \phi^2} +e^{-b \phi}
\frac {\partial^2\psi} {\partial \sigma^2} \right) +
3 v_k^2 a^2\left (k-\frac{1}{6} m^2 a^2
e^{2 b \phi} \sigma^2\right) \psi=0,
\end {equation}
where $p $ is factor ordering which introduction is caused
by ambiguity of commutation properties of the operators $a$ and $p_a$. Transformation
of wave function
$$
\psi\rightarrow a ^ {-p/2} \psi
$$
allows us to exclude the first derivative in Eq. (\ref {wd0}) that gives:
\begin {equation}
\label {wd}
-\frac{1}{12}\frac {\partial^2\psi} {\partial a^2} + \frac {1} {2 a^2}
\left (\frac {\partial^2\psi} {\partial \phi^2} +e ^ {-b \phi}
\frac {\partial^2\psi} {\partial \sigma^2} \right) +
\left [-\frac {p} {24 a^2} \left (1-\frac {p} {2} \right) +
3 v_k^2 a^2\left (k-\frac{1}{6} m^2 a^2
e^{2 b \phi} \sigma^2\right) \right] \psi=0.
\end {equation}
Let us choose the factor ordering $p=1$. Also we shall use the following
redefinitions: $ \phi, \sigma \rightarrow \sqrt {6} \phi, \sigma $,
$b \rightarrow \sqrt {1/6} b $, $v_k \rightarrow 1/6 v_k$. Finally, from Eq. (\ref {wd}) we have:
\begin {equation}
\label {wd1}
-\frac {\partial^2\psi} {\partial a^2} + \frac {1} {a^2}
\left (\frac {\partial^2\psi} {\partial \phi^2} +e ^ {-b \phi}
\frac {\partial^2\psi} {\partial \sigma^2} \right) +W \psi=0,
\end {equation}
where superpotential
\begin {equation}
\label {poten}
W =-\frac {1} {4 a^2} +
 v_{k}^{2} a^2\left (k-a^2 e ^ {2 b \phi} m^2 \sigma^2 \right).
\end {equation}

\section {Problem statement}

The obtained equation describes quantum evolution of the universe. Thus
the statement of problem essentially depends on type of a model
of the universe which is examined: closed, flat or open. In view of equality to zero of full
energy of the universe, it is clear from form of
superpotential (\ref {poten}) that consideration of the problem is reduced to research either
the process of tunneling in case of the closed model or the process
of above the barrier reflection of the universe's wave function in case of flat
and open models. Such consideration was carried out repeatedly in
various statements. However, in most cases one-dimensional problems were examined.
In this cases wave function depends only on one variable -
scale factor. At that it was possible to obtain exact
solutions in a number of problems \cite {ref:Wilt}. And, that is even more important, in one-dimensional statement it is seems
possible to find clear physical sense of the wave function. In a case of
multidimensional problems, when the wave function depends also from
field variables, the problem is complicated not only by
mathematical difficulties but also by complexities with understanding of
physical sense of the multidimensional wave function. In this connection some
authors proposed the methods of reducing of multidimensional problems to one-dimensional one.
It allows to give to the wave function physically clear sense \cite {ref:Most}.

In the given work we shall use one of such ways which was suggested
in Ref. \cite {ref:Gur1}. The essence of this method consists in consideration of semi-classical
evolution of a model of the universe. In this case the corresponding
Hamilton-Jacobi equation, similar to the equation from
classical mechanics, is written down for the Wheeler-DeWitt equation. This equation is nonlinear first-order
partial differential equation. For its solution
the known method is used. The essence of this method consists in representation
of the given nonlinear partial differential equation
through a system of the first-order ordinary differential equations
(so-called set of characteristic equations). The given equations
describe a set of characteristics which are form an integral surface
of initial  partial differential equation. Distinctive feature of
the system of characteristic equations is that the functions, entering
in equations, depend only on one argument. Thus, initially
multidimensional problem is reduced to one-dimensional one
that allows to find physically sensible one-dimensional wave
functions along each of characteristics. Then for calculation, for example, of
permeability of a barrier, it is necessary only to sum and to average
permeability on all characteristics.

\section {Derivation and solving of a set of characteristics}

Let us pass directly to research of our model.

\subsection*{Closed model}

At the beginning, let us consider the
case of closed universe. As it was already specified above, in this case the process of
tunneling of the wave function will be take place. In semi-classical treatment
of the given problem we can present the wave function as $\psi=e^{-S}$.
Then, being limited to zero approximation, we obtain
the following equation for action $S$ from Eq. (\ref {wd1}):
\begin {equation}
-\left(\frac{\partial S}{\partial a} \right)^2 +\frac{1}{a^2}\left[
\left(\frac{\partial S}{\partial \phi}\right)^2 +
e^{-b\phi}\left(\frac{\partial S}{\partial \sigma}\right)^2
\right] +W=0.
\end {equation}
The corresponding set of characteristics under a barrier will be (see.,
e.g., Ref. \cite {ref:Kamke}):
\begin {eqnarray}
\label {char}
\frac {d\phi}{d a} &=&-\frac {q} {a^2 F}, \quad
\frac {d\sigma}{d a} =-\frac {e ^ {-b \phi} v} {a^2 F}, \nonumber \\
\frac {d q} {d a} &=&-\frac {1} {2} \frac {b e ^ {-b \phi} v^2} {a^2 F}-
\frac {b v_{k}^{2} a^4 e^{2 b \phi} m^2 \sigma^2} {F}, \quad
\frac {d v} {d a} =-\frac {v_{k} ^ {2} a^4 e ^ {2 b \phi} m^2 \sigma} {F}, \\
\frac {d S} {d a} &=& \frac {W} {F}, \nonumber
\end {eqnarray}
where $q =\partial S / \partial \phi $ and
 $v =\partial S / \partial \sigma $ are the generalized momenta and
 $F =\sqrt {(q^2+v^2)/a^2+W} $.

This equations describe a process of tunneling of a "particle" along
the characteristics. Each of such characteristics differs from another
by definition of corresponding initial conditions. For this purpose let us analyse
the potential (\ref {poten}) at $k=1$. As we see, the whole minisuperspace
is divided into three areas (see Fig. (\ref{@w_cl})):
\FIGURE[!h]{\epsfig{file=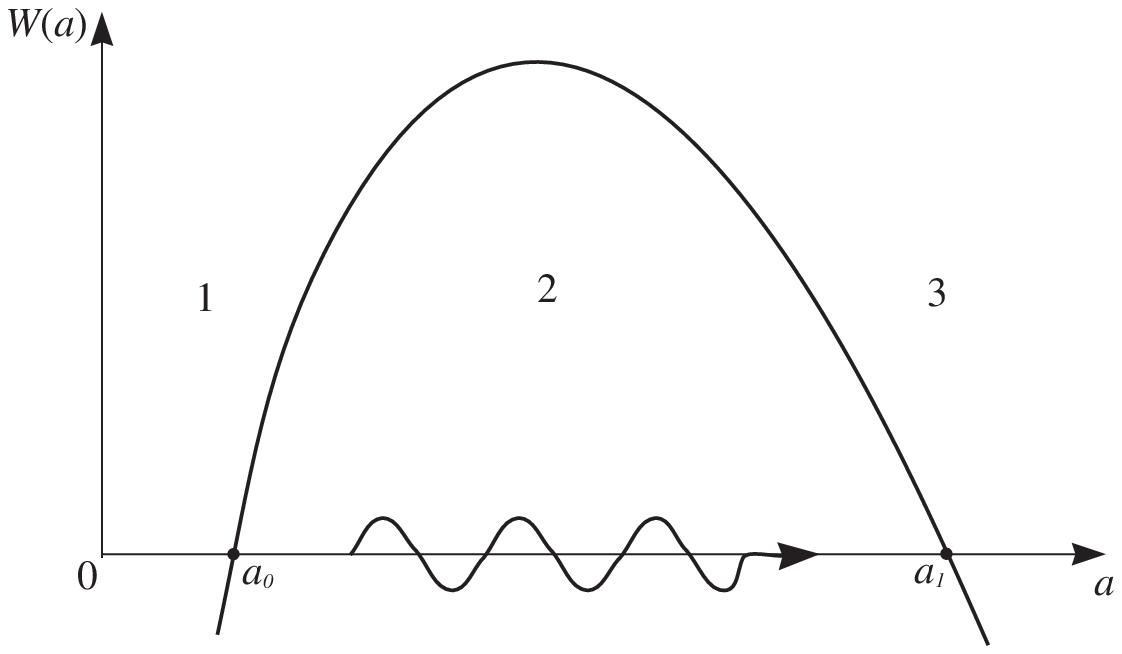,width=.57\columnwidth} \qquad
\caption{The conventional form of an one-dimensional barrier is shown.
Tunneling of the wave function occurs from the point $a_0$, where initial conditions are set,
into the point of an output from under the barrier $a_1$. The initial
conditions in $a_0$ define the
form of a barrier which is individual for each characteristic.} \label{@w_cl}}
1) internal area, where there is no
classical space-time and the metrics is under the strong
quantum fluctuations; 2) classically forbidden area under a barrier;
3) classically allowed area. Initial conditions are set on bound
between internal and classically forbidden areas. For finding of the bound,
it is necessary to equate superpotential (\ref {poten}) to zero. In generally,
it will be a surface. However, if to choose as initial conditions
$ \phi, \sigma=0 $, evolution of the model will begin in a point
$a_0 = (1/4 v_{1} ^2) ^ {1/4} $, where $v_1$ is a three-volume of the closed universe. In
case of sphere $v_1=2\pi^2$. Further, we should set initial momenta
$q=q_0$ and $v=v_0$ which will vary from 0 up to $\infty$. So,
all characteristics will be differ by different values
of momenta. Using the specified initial conditions and solving numerically
the set of equations (\ref {char}) we have obtained the following results
(see Figs. (\ref{@f_b})).
\FIGURE[!h]{\epsfig{file=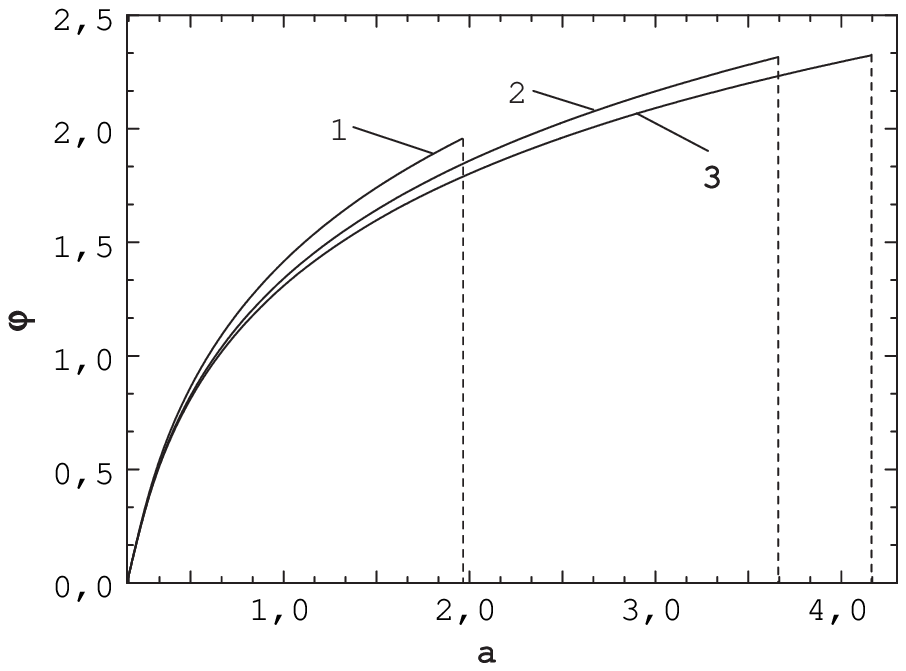,width=.47\columnwidth} \qquad
\epsfig{file=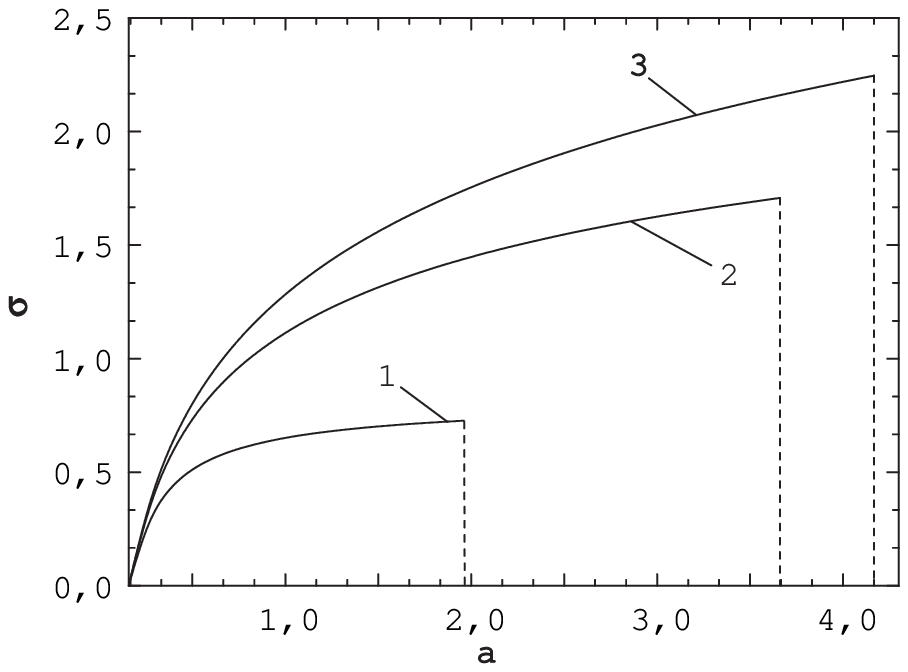,width=.47\columnwidth} \caption{The dependence $\phi(a)$ (left panel) and
$\sigma (a)$ (right panel) for
different $b$: 1 - for $b=1$, 2 - for $b=0.2$, 3 - for $b=0.02$.
Dotted lines are pointing out the points of output from under a barrier.} \label{@f_b}}

This implyies that in process of under a barrier
evolution, there is a growth both inflaton and dilaton fields.
Thus it is important to note that thickness of a barrier and, consequently,
the rate of growth of fields strongly depends on parameter $b$:
the less value of $b$ means the more width of a barrier and the more value of fields on an output
from under a barrier.
On the other hand it is clear that after an output from under a barrier the parameters of
the model should provide conditions for the subsequent
inflationary stage. At least, there are two such conditions: 1) inflaton
field should be the Planck's order; 2) it should
varies slowly enough to ensure long
inflationary stage. In the given model such conditions are provided
in a case when the coupling constant $b$ is small enough.
This conclusion is in agreement
with specified in Ref. \cite {ref:Bous} restriction on value of $b<<1$.

\subsection*{Flat model}

It is possible also to consider evolution of the given model for
cases when $k=0,-1$. It is obvious that in these cases the potential barrier will be
lay in negative area and only above the barrier reflection is possible.
In view of absence of principle distinctions of process of above the barrier reflection
for open and flat model, we shall consider, for example, an evolution of flat
model (i.e. we should put in Eq. (\ref {poten}) $k=0$). Since full energy of the universe
is equal to zero and the superpotential always is less than zero, then in case of flat model
evolution of the universe will always take place in classically allowed area.
In this connection let us search  for solution of equation (\ref {wd}) as $\psi=e^{-i S}$.
Then we obtain:
\begin {equation}
-\left (\frac {\partial S} {\partial a} \right) ^2 +\frac {1} {a^2} \left [
\left (\frac {\partial S} {\partial \phi} \right) ^2 +
e ^ {-b\phi} \left (\frac {\partial S} {\partial \sigma} \right) ^2
\right]-W=0.
\end {equation}
Then the set of characteristics reads:
\begin {eqnarray}
\label {char2}
\frac {d \phi} {d a} &=&-\frac {q} {a^2 F}, \quad
\frac {d \sigma} {d a} =-\frac {e ^ {-b \phi} v} {a^2 F}, \nonumber \\
\frac {d q} {d a} &=&-\frac {1} {2} \frac {b e ^ {-b \phi} v^2} {a^2 F} +
\frac {b a^4 e ^ {2 b \phi} m^2 \sigma^2} {F}, \quad
\frac {d v} {d a} = \frac {a^4 e ^ {2 b \phi} m^2 \sigma} {F}, \\
\frac {d S} {d a} &=&-\frac {W} {F}, \nonumber
\end {eqnarray}
where $F =\sqrt{(q^2+v^2)/a^2-W}$. (Here we have choose $v_0=1$ (see Ref. \cite {ref:Coule}).) For numerical research
of this system, it is necessary
to set the certain boundary conditions. It can be made as follows.
Two conditions, necessary for the beginning of inflationary stage, were specified above.
 First of these conditions can be written as
the following restriction on value of inflaton and dilaton fields
in the beginning of inflation \cite {ref:Bous}:
\begin {equation}
\label {constr}
e ^ {2b\phi} m^2\sigma^2=1.
\end {equation}
The fulfilment of the second condition can be checked directly at
numerical research of the set of equations (\ref {char2}).
Thus evolution of the universe will pass in two stages: first, the stage of contraction will
occurs from a point of the beginning of inflation
up to a turn point. Then, after the bouncing off,
the model will come on a stage of expansion.

So,
having use the condition (\ref {constr}) and, as well as in a case of closed
model, setting some values of momenta $q=q_0$ and $v=v_0$ in a point
of the beginning of inflation we can carry out numerical research of the set of characteristics.
Its results are presented in Figs. (\ref{@f_b_1}).
\FIGURE[!h]{\epsfig{file=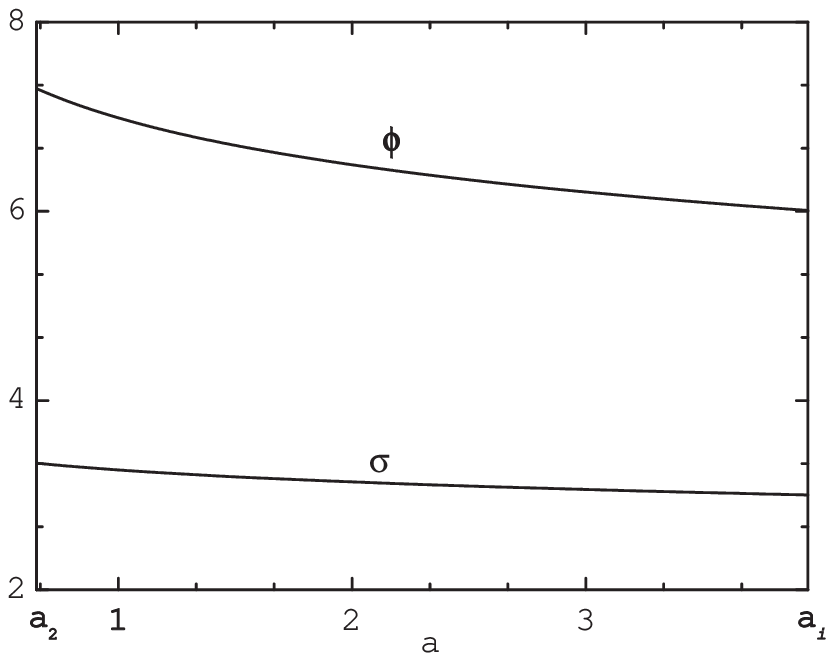,width=.47\columnwidth} \qquad
\epsfig{file=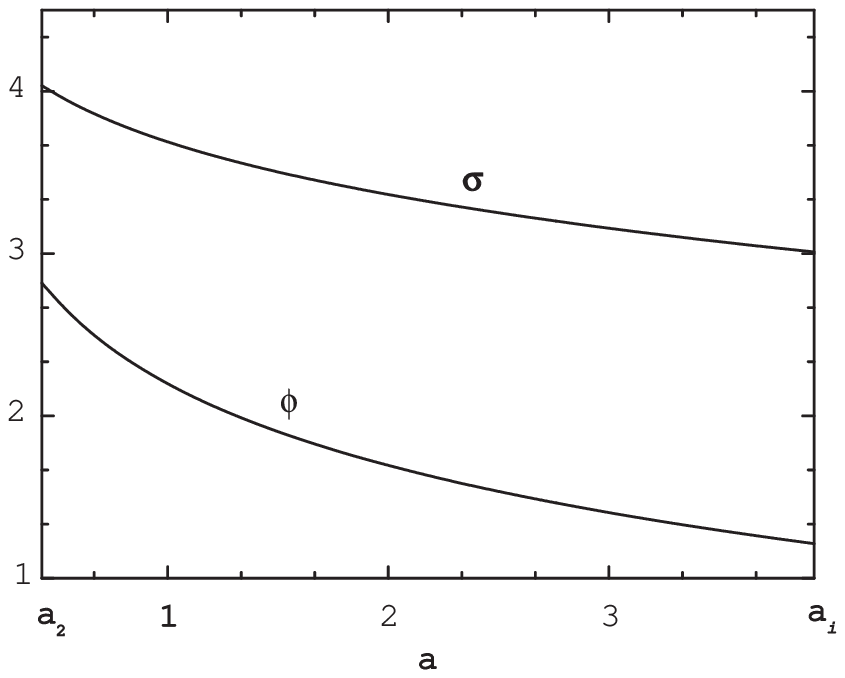,width=.47\columnwidth} \caption{The dependence $\phi(a)$ and
$\sigma (a)$  for
$b=0.2$ (left panel) and for $b=1$ (right panel). Here $a_2$ is a point of above the barrier
reflection, $a_i$ - point of the beginning of inflation.
} \label{@f_b_1}}

\section {Probabilities of tunneling and above the barrier reflection of
wave function}

In the considered above models,  evolution of scalar fields was investigated
during tunneling and above the barrier reflection. However, it
is also interesting to find probabilities of these processes. It appears
that it is possible to make in analytical form because of one feature of
the considered sets of characteristics. It is clear from
the numerical analysis of Eqs. (\ref {char}) in case of tunneling  that
action $S $ is large at small values of momenta $q$ and $v$.
Thus permeability of a barrier will be exponentially
small in accordance with the formula
\begin {equation}
\label {pntr}
D = \exp (-2 S).
\end {equation}
Therefore the basic contribution to permeability will be brought by characteristics
with large momenta for which the action is small. In this case
it is possible to simplify essentially the set of characteristics (\ref {char})
because at large momenta's values their practically
does not vary during tunneling \cite {ref:Gur1, ref:Fol}.
Then Eqs. (\ref {char}) will become:
\begin {eqnarray}
\label {char3}
\frac {d \phi} {d a} &=&-\frac {q} {a^2 F}, \quad
\frac {d \sigma} {d a} =-\frac {e ^ {-b \phi} v} {a^2 F}, \nonumber \\
q&=&const, \quad
v=const, \\
\frac {d S} {d a} &=& \frac {W} {F}, \nonumber
\end {eqnarray}
where $q, v>> 1 $, $F\approx A/a $ and $A^2=q^2+v^2 $. Then it is easy to find
\begin {equation}
\label {fld1}
\phi =\frac {q} {A} \ln {\frac {a_0} {a}}, \quad
\sigma =-\frac {v} {b q} \left (\frac {a} {a_0} \right) ^ {b q/A}
\end {equation}
and
\begin {equation}
\label {act1}
S = \frac {1} {A} \left [-\frac {1} {4} \ln {\frac {a_ *} {a_0}} +v_{1} ^ {2} \frac {1} {4}
(a _ {*} ^ {4}-a _ {0} ^ {4})-\frac {m^2 v^2} {b^2 q^2} v_{1} ^ {2} \frac {1} {6}
(a _ {*} ^ {6}-a _ {0} ^ {6}) \right].
\end {equation}
In these expressions $a_0 $ is a point of an entrance of wave function under a barrier,
$a_*$ is a point of an output. Last one is determined from the condition
$W(a _{*},\phi_{*},\sigma_{*}) = 0$ and, finally, is a
function of momenta $q$ and $v$.

As it has already been mentioned above, these formulas
are true only at large values of $q$ and $v$. However, in this case action $S$
is small that does not allow to use
the usual formula for permeability of a barrier (\ref {pntr})
which is applicable only at large $S$. Therefore in this case
it is necessary to use the formula of generalized WKB \cite {ref:Froman}:
\begin {equation}
\label {penet}
D = \frac {\exp (-2 S)} {1+\exp (-2 S)},
\end {equation}
which is used at any $S$. Then, using Eqs. (\ref {act1}),
(\ref {penet}) and averaging on all characteristics, we can obtain the following
expression for permeability of a barrier:
\begin {equation}
\label {penet2}
D = \lim _ {\overline {q}, \overline {v} \rightarrow \infty}
\frac {1} {\overline {q} \, \overline {v} \,} \int\limits _ {0} ^ {%
\overline {q} \,} \int\limits _ {0} ^ {\overline {v} \,} %
dqdv\frac {\exp \left (-2S\left (q, v\right) \right)} {1 +\exp \left (
-2S\left (q, v\right) \right)}.
\end {equation}

\FIGURE[!h]{\epsfig{file=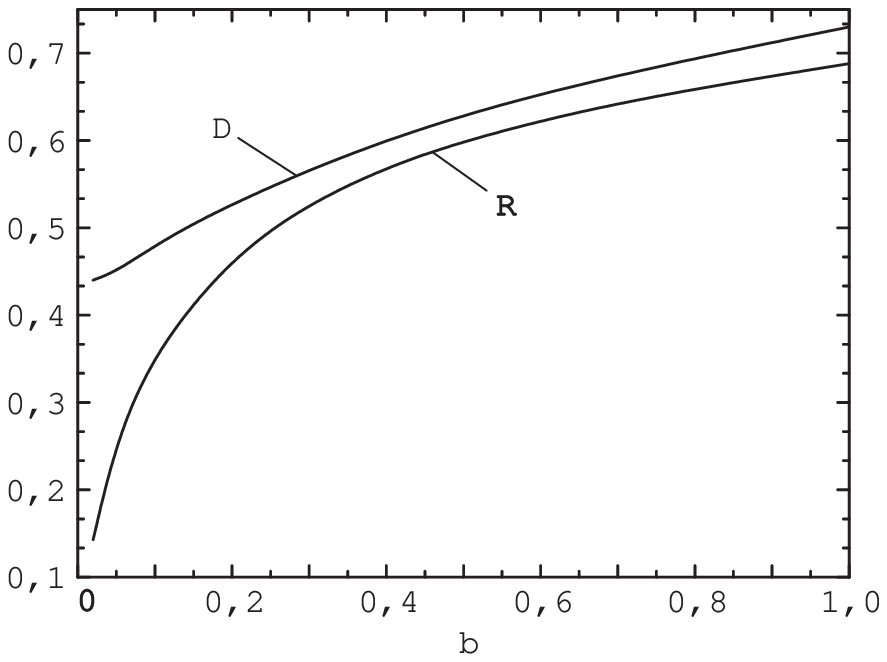,width=.47\columnwidth}
\caption{The dependence of permeability $D$ and reflectivity $R$ from $b$.
} \label{@R_D_b}}
In a case of flat model the above the barrier reflection will occurs
near to a maximum of superpotential (\ref {poten}) where it is small on absolute
value in comparison with the term $(q^2+v^2)/a^2$ from Eqs. (\ref {char2}). Then
we shall have again the same expressions (\ref {fld1}) for fields.
For calculation of reflectivity we use
the known formula \cite{ref:Land}:
\begin {equation}
\label {refl}
R = \exp \left (-4 \, {\rm Im} \int _ {a _ {1}} ^ {a _ {2}} da \sqrt{-W(a)}
 \right),
\end {equation}
where $a_1$ is an arbitrary point on a real axis and $a_2$ is
a turn point calculated from the condition $W=0$. In our case, using
Eqs. (\ref {fld1}) and (\ref {poten}) for $k=0 $, we have:
\begin {equation}
W =-\frac {1} {4 a^2}-\frac {m^2 v^2} {b^2 q^2} a^4
\end {equation}
and
\begin {equation}
a_2 =\left (\frac {b q} {2 m v} \right) ^ {1/3} \exp {\left[i\pi (2n+1)/6\right]}.
\end {equation}
Then the reflectivity, averaged on all momenta, will be:
\begin {equation}
\label {refl2}
R = \lim _ {\overline {q}, \overline {v} \rightarrow \infty}
\frac {1} {\overline {q} \, \overline {v} \,} \int\limits _ {0} ^ {%
\overline {q} \,} \int\limits _ {0} ^ {\overline {v} \,} %
dqdv \exp \left (-\frac {\pi} {3} \sqrt {\frac {m} {b} \frac {v} {q}} \right).
\end {equation}

The obtained formulas (\ref {penet2}) and (\ref {refl2}) allow us to find
the probabilities at various values of parameter $b$ (Fig. \ref {@R_D_b}).
As we see  from figure, in case of the closed model permeability is the larger when
$b$ is larger (i.e. when the barrier is more thin), that is quite natural. On the other hand,
as was noted above, than $b$ is smaller, the
inflaton field on an output from under a barrier is larger and the conditions for inflation
are better. So, certain compromise is necessary at a choice of parameter $b$
for maintenance both enough long inflation and
high probability of tunneling.

In a case of flat model the similar situation occurs. Here, then $b$ is larger, decreasing of the inflaton field $\sigma$
is faster on the expansion stage (see Figs. (\ref{@f_b_1})). Then the conditions for inflation at identical
values of the inflaton field in the point of above the barrier reflection $a_2$ in a case of
large $b$ will be worse than at small $b$. At that, the probability of above the barrier
reflection is increasing at large $b$, as in a case of closed model.

\bigskip
I am grateful to H.-J. Schmidt and H. Kleinert for discussions. This work was supported by ISTC project KR-677.

\end{document}